\documentclass[prb,twocolumn,amsmath,amssymb,amsfonts,floatfix,superscriptaddress]{revtex4}

\usepackage{graphicx}
\usepackage{color}
\usepackage[utf8]{inputenc}
\usepackage{dcolumn}
\usepackage{hyperref}
\usepackage{comment}
\usepackage[caption=false]{subfig}
\usepackage[english]{babel}
\usepackage{indentfirst}
\hypersetup{colorlinks=true,breaklinks,linkcolor=blue,urlcolor=blue,citecolor=blue}

\newcommand{\DF}{\texttt{dualfermion}}
\newcommand{\triqs}{\textsc{triqs}}
\newcommand{\cthyb}{\texttt{cthyb}}

\newcommand{\up}{\uparrow}
\newcommand{\dn}{\downarrow}
\newcommand{\Rv}{\ensuremath{\mathbf{R}}}
\newcommand{\kv}{\ensuremath{\mathbf{k}}}

\newcommand{\av}[1]{\ensuremath{\left\langle #1 \right\rangle}}

\renewcommand{\Re}{\operatorname{Re}}
\renewcommand{\Im}{\operatorname{Im}}

\usepackage{tikz}
\usetikzlibrary{shapes.misc}
\usetikzlibrary{arrows.meta}

\definecolor{mycolor1}{RGB}{228,26,28}
\definecolor{mycolor2}{RGB}{55,126,184}
\definecolor{mycolor3}{RGB}{77,175,74}
\definecolor{mycolor4}{RGB}{152,78,163}
\definecolor{mycolor5}{RGB}{255,127,0}

\definecolor{mylight1}{RGB}{127,201,127}
\definecolor{mylight2}{RGB}{253,192,134}

\begin{document}

\title{Second-order dual fermion for multi-orbital systems}

\author{Erik G. C. P. van Loon}
\affiliation{Institute for Theoretical Physics and Bremen Center For Computational Materials Science, University of Bremen, Bremen, Germany}

\begin{abstract}
 In dynamical mean-field theory, the correlations between electrons are assumed to be purely local. 
 The dual fermion approach provides a systematic way of adding non-local corrections to the dynamical mean-field theory starting point.
 Initial applications of this method were largely restricted to the single-orbital Hubbard model. 
 Here, we present an implementation of second-order dual fermion for general multi-orbital systems and use this approach to investigate spatial correlations in SrVO$_3$.
 In addition, the approach is benchmarked in several exactly solvable small systems.
\end{abstract}

\maketitle

Diagrammatic extensions of dynamical-mean field theory~\cite{Rohringer18} are an important family of computational approach for the study of correlated electron systems. The mean-field description of local dynamical correlations~\cite{Georges96} forms the starting point for a (perturbative) calculation of spatial correlations. Since spatial correlations vanish in the opposite limits of small and large interaction~\cite{Rubtsov08,Rubtsov09} of the Hubbard model, these approaches provide a reasonable interpolation of correlated electron physics across large parts of the phase diagram.

An important example of this class of methods is the dual fermion~\cite{Rubtsov08} approach. 
The dual fermion~\cite{Rubtsov08,Rubtsov09,Hafermannphd} technique is based on a transformation to new degrees of freedom with (hopefully) simpler and numerically more tractable correlations. 
These degrees of freedom are subsequently treated with a further theoretical technique (e.g., perturbation theory, ladder resummations~\cite{Hafermann09}, parquet~\cite{Krien20,Krien20b}, renormalization group approaches~\cite{Wentzell15} or diagrammatic Monte Carlo~\cite{Iskakov16,Gukelberger17}). The dual fermion approach has been used to study the effect of magnetic fluctuations on the electronic properties~\cite{Rubtsov08,Rubtsov09,Hirschmeier15,vanLoon18,Schafer20}, the metal-insulator transition~\cite{vanLoon18d,Tanaka19}, susceptibilities~\cite{Brener08}, critical exponents~\cite{Antipov14}, weak localization~\cite{Astleithner20}, disordered systems~\cite{Terletska13,Yang14,Haase17} and the competition of charge, spin and superconducting fluctuations~\cite{Otsuki14,Astretsov20} 
These applications largely occured in systems with either a single correlated orbital per unit cell or where all orbitals are equivalent~\cite{Hafermann08,Iskakov18,Hirschmeier18}. 

This work considers the second-order dual perturbation theory in multi-orbital systems. This approach is simple, computationally affordable and is a good indicator of the relevance of spatial correlations~\cite{vanLoon18d}.
Whereas ladder approaches require the choice of a particular set of fluctuation channels (e.g., particle-hole or particle-particle), in second order dual fermion all processes are included at the same level of sophistication and the method is unbiased. 
Although the second-order dual fermion has also seen applications to inhomogeneous systems~\cite{Takemori16,Takemori18}, this work is restricted to periodic systems.

The implementation is based on the \triqs~\cite{triqs} toolkit (version 3.0) and can be used together with the \triqs\ continuous-time hybridization (\cthyb) solver~\cite{cthyb} as well as with the exact diagonalization solver pomerol~\cite{pomerol} (via pomerol2triqs).
A convenient interface to these other \triqs\ applications is provided. 
The source code is publicly available\footnote{Currently hosted at \url{https://github.com/egcpvanloon/dualfermion/}, installation instructions and examples are available there.}. 

This implementation (\DF) aims to be applicable to a general set of periodic correlated electron models. In particular, neither spin symmetry nor lattice symmetries are assumed. This is considered an acceptable trade-off since the usual bottleneck of dual fermion calculations lies in the solution of the auxiliary impurity model, not in the dual perturbation theory. The code does use the ``block structure'' of \triqs\ and equivalent blocks can be exploited to achieve some speed-up. E.g., in the paramagnetic Hubbard model the self-energy only needs to be calculated for the spin $\up$ block. 

The structure of this manuscript is as follows:
It starts with an overview of the method in Sec.~\ref{sec:method}, including relevant details for the multi-orbital implementation. In Sec.~\ref{sec:srvo3}, the method is applied to SrVO$_3$ and the results are compared to dynamical vertex approximation (D$\Gamma$A~\cite{Toschi07,Galler19}) results available in the literature. In Sec.~\ref{sec:clustertesting}, several examples of small, exactly solvable models are provided. These serve as useful tests of the implementation, since second-order dual fermion is guaranteed to become exact in several limits. We track how deviations between the second-order dual perturbation theory and the exact solution start to appear as one moves away from these limits.
These examples illustrate some of the potential applications of multi-orbital dual fermion. In addition, the approach could also prove useful for the study of inhomogeneous bilayers containing both weakly and strongly correlated electrons~\cite{Chen20}.

\section{Method and Implementation}
\label{sec:method}

\begin{figure*}
\begin{tikzpicture}[node distance=1.7cm,
            base/.style = {rectangle, rounded corners, draw=black,
                           minimum width=4cm, minimum height=1cm,
                           text centered, font=\sffamily},
          python/.style = {base,fill=mylight1},
             ext/.style = {base,fill=mylight2},
]

\node at (-2.5,0.5) {(a)};

 \node (constr) [python]            {\texttt{dualfermion.Dpt()}};
 \node (load)   [python,
                below of=constr]    {
                                        \begin{tabular}{c}
                                        \texttt{Delta << ...}\\ 
                                        \texttt{gimp << ...}\\ 
                                        \texttt{G2\_iw << ...}\\
                                        \end{tabular}
                                        };
 \node (solver) [ext,
                xshift=4cm,
                right of=load]      {
                                        \cthyb, pomerol, \ldots
                                    };
 \node (run)    [python,
                below of=load
                ]
                                    {\texttt{run()}};
 \node (out)    [ext,
                below of=run
                ]
                                    {output};
 \node (hk)    [ext,
                xshift=4cm,
                right of=constr
                ]
                                    {$H_\kv$, $\beta$, freq. mesh};

 \draw[-latex] (out) -| 
    node[ext,pos=0.167,
            minimum width=0cm,
            minimum height=0cm,
            circle] 
    {$\Delta$}            
    node[ext,pos=0.333,
            minimum width=0cm,
            minimum height=0cm,
            circle] 
    {$G_0$}
    (solver) ;
 \draw[-latex] (solver) -- (load) ;
 \draw[-latex] (load) -- (run) ;
 \draw[-latex] (constr) -- (load) ;
 \draw[-latex] (hk) -- (constr) ;
 \draw[-latex] (run) -- (out);

 \node[anchor=south] at (constr.north) {python workflow};
 \node[anchor=south] at (hk.north) {input};
 
\end{tikzpicture}\hspace{1.5cm}
\begin{tikzpicture}[node distance=1.5cm,
            base/.style = {rectangle, rounded corners, draw=black,
                           minimum width=2.5cm, minimum height=1cm,
                           text centered, font=\sffamily},
]

\node at (-2,0.5) {(b)};

\node (gd0k)
        [base,fill=mycolor1!50]
        {$\tilde{G}_0(i\omega,\kv)$};
        
\node (gd0r)
        [base,fill=mycolor1!50,
        below of=gd0k]
        {$\tilde{G}_0(i\omega,\Rv)$};
 
\node (sigmad0r)
        [base,fill=mycolor1!50,
        below of=gd0r]
        {$\tilde{\Sigma}_0(i\omega,\Rv)$};

\node (sigmad0k)
        [base,fill=mycolor1!50,
        below of=sigmad0r]
        {$\tilde{\Sigma}_0(i\omega,\kv)$};

\node (gk)
        [base,fill=mycolor1!50,
        below of=sigmad0k]
        {$G(i\omega,\kv)$};
        
\node [anchor=south] at (gd0k.north) {\texttt{run()}};        
\draw [->] (gd0k) -- node[left] {FT} (gd0r); 
\draw [->] (gd0r) -- node[left] {diagrams} (sigmad0r); 
\draw [->] (sigmad0r) -- node[left] {FT} (sigmad0k); 
\draw [->] (sigmad0k) -- node[left] {inv. dual} (gk);

\end{tikzpicture}

\caption{Structure of a dual fermion calculation. (a) Overview of the computational scheme as exposed to python.
(b) Internal structure of \texttt{run()}, implemented in \texttt{C++}.
}
\label{fig:flowchart}
\end{figure*}
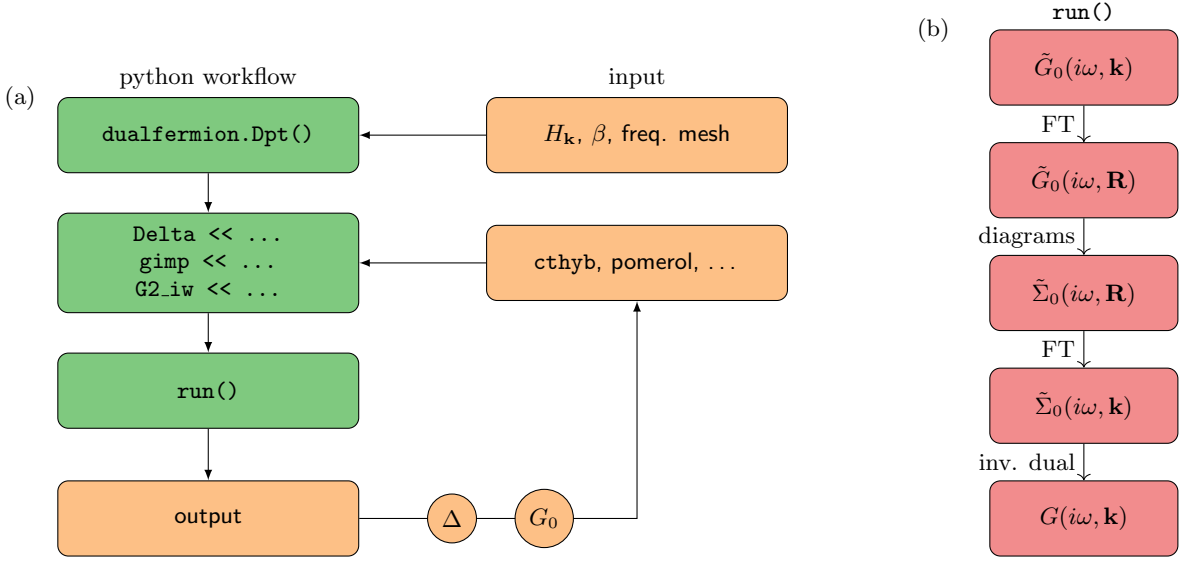

The formal derivation and central ideas of the dual fermion approach can be found in the literature~\cite{Rubtsov08,Hafermannphd}, here we provide a brief overview of the necessary equations and how they are implemented in \DF. 

We consider lattice Hamiltonians of the form
\begin{align}
 H =& -\sum_{\Rv_i,\Rv_j} \sum_{\sigma,\alpha\beta} h^\sigma_{\alpha\beta,\Rv_i-\Rv_j} c^\dagger_{\sigma,\alpha,\Rv_i}c^{\phantom{\dagger}}_{\sigma,\beta,\Rv_j}
 \notag \\
 &+\sum_\Rv U^\text{site}[c^\dagger_{\Rv},c^{\phantom{\dagger}}_{\Rv}].
\end{align}
Here $(\sigma,\alpha)$ and $(\sigma,\beta)$ are orbital labels. $\Rv$ labels the unit cells (atoms, clusters, etc.), $c^\dagger_{\alpha,\Rv_i}$ is the creation operator for a fermion in orbital $\alpha$ in cell $\Rv_i$, $h$ is the inter-cell hopping matrix and $U^\text{site}$ is the intra-cell Hamiltonian, which is allowed to contain arbitrary interactions. Since all intra-cell terms are included in $U^\text{site}$, $h_{\Rv_i-\Rv_j}=0$ for $\Rv_i=\Rv_j$. 
We denote the Fourier transform of $h$ as $H(\kv)$, where the orbital matrix structure is implied.

The orbital label $\sigma$ is special because $h$ is diagonal in this label and this simplifies the structure of all equations. We will call $\sigma$ the block label. An example of such a block structure is the spin in the paramagnetic Hubbard model. For the block structure to hold, we also require that $U^\text{site}$ conserves $\sigma$.

In the dynamical mean-field theory spirit, the local correlations are initially approximated by a single-cell reference model, the so-called auxiliary impurity model, with action
\begin{align}
  S =& -\sum_{\sigma,\alpha\beta} (i\nu_n+\mu -\Delta^{\sigma}_{\nu_n,\alpha\beta}) c^\dagger_{\sigma,\alpha,\nu_n}c^{\phantom{\dagger}}_{\sigma,\beta,\nu_n}
 +U^\text{site}[c^\dagger,c^{\phantom{\dagger}}].\label{eq:action}
\end{align}
Here, $\nu_n$ denotes a Matsubara frequency and $\Delta^\sigma_{\nu_n,\alpha\beta}$ is the hybridization function.
The idea of dynamical mean-field based approaches is that, given some $\Delta$, this impurity model can be solved numerically and all relevant correlation functions can be calculated. 

In particular, we obtain the single-particle Matsubara Green's function $g^\sigma_{\alpha\beta}(i\nu_n)$,
which is diagonal in the block label $\sigma$ and a matrix in the orbitals $\alpha$,$\beta$. 
Similarly, the connected two-particle Green's function $\gamma$ has two block and four orbital labels and is defined as 
\begin{align}
 \gamma^{\sigma_1\sigma_2}_{\alpha\beta\gamma\delta}(w, n_1, n_2) 
            = 
            &(G^2)^{\sigma_1\sigma_2}_{\alpha\beta\gamma\delta}(w, n_1, n_2) \notag \\
            &
            +\beta\delta^{\sigma_1\sigma_2}\delta(n_1,n_2)g^{\sigma_1}_{\delta,\alpha}(n_1)g^{\sigma_2}_{\beta,\gamma}(n_1+w) \notag \\
            &
            - \beta \delta_w g^{\sigma_1}_{\beta\alpha}(n_1)g^{\sigma_2}_{\delta\gamma}(n_2).\end{align}
where $G^2$ is the two-particle Green's function that is obtained from the impurity solver, $n_1$ and $n_2$ are fermionic Matsubara frequencies and $w$ is a bosonic Matsubara frequency.

From the impurity quantities, we construct the (bare) dual Green's function $\tilde{G}$, which is block diagonal and a matrix in orbital space and is defined in momentum space as
\begin{align}
 \tilde{G}^{\sigma}&(n_1,\kv) 
 \label{eq:dual}\\
 =&  \left[ g^{\sigma}(n_1) + g^{\sigma}(n_1)\left(\Delta^{\sigma}(n_1) - H^{\sigma}(\kv)   \right)g^{\sigma}(n_1)\right]^{-1} \notag \\
 &- \left[g^{\sigma}(n_1)\right]^{-1},\notag 
\end{align}
where $^{-1}$ stands for matrix inversion. 

From the dual Green's function and the vertex, the dual self-energy $\tilde{\Sigma}$ is calculated. The explicit self-energy diagrams will be given in Sec.~\ref{sec:diagrams}. Finally, the lattice Green's function is calculated from this dual self-energy as
\begin{align}
 G^{\sigma}(n_1,\kv) = ( (g^\sigma(n_1)+\tilde{\Sigma}^\sigma(n_1,\kv))^{-1} + \Delta^\sigma(n_1) - H^\sigma(\kv)  )^{-1}. \label{eq:back}
\end{align}

At this point, we should note that this implementation differs from most of the existing dual fermion literature in the normalization of the Hubbard-Stratonovich decoupling~\cite{Rubtsov08,Hafermannphd}. 
This is done here by using $\alpha_\text{HS}=1$ instead of the usual choice of $\alpha_\text{HS}=g^{-1}$. 
This leads to slightly different relations between dual and original fermions, Eqs.~\eqref{eq:dual} and \eqref{eq:back}, but avoids unnecessary (matrix) inversions. Furthermore, divisions by $\Delta_\nu-t_\kv$ are avoided in the present implementation to ensure that the atomic limit $t=0$, $\Delta=0$ is well-defined.

\subsection{Computational flow}

Figure~\ref{fig:flowchart}(a) gives a graphical overview of the program. It consists of the following steps:
\begin{enumerate}
 \item Construction of the dual perturbation theory object using \texttt{dualfermion.Dpt()}. The desired orbital, lattice and frequency structure are passed as arguments to the constructor.
 \item Assigning properties of the auxiliary impurity model to the dual perturbation theory object using the \triqs\ \texttt{<<} notation. In particular, the hybridization function (which was the input of the impurity model) and the one- and two-particle Green's function (output of the impurity model).
 \item Running the dual pertubation theory, using \texttt{run()}, eventually calculating $G(\nu,\kv)$. The structure of this part of the program is shown in more detail in Fig.~\ref{fig:flowchart}(b). The dual perturbation theory is entirely implemented in C\texttt{++} and the end user only needs to call the function \texttt{run()} from python.
 \item The resulting output can then be used to generate a new impurity model, so called ``outer self-consistency''. With \texttt{calculate\_sigma=False}, this is equivalent to the usual DMFT self-consistency loop.
\end{enumerate}

Input to the code is given in three stages, as is visible in Fig.~\ref{fig:flowchart}(a). In the first step (the constructor), a python object that will be used throughout the calculation is initialized. In this first step, arrays are constructed that will hold data later, so it is necessary to know the size of all relevant objects. In the second step, the Green's function objects that are used in the calculation are loaded. In other words, the actual physical data is transferred here. Finally, parameters that control the execution of the dual perturbation theory are passed via the run command. 

In a self-consistency loop, the first step only needs to be done once per calculation, whereas the second and third step happen in every iteration.
For those familiar with \cthyb, these three steps correspond to \texttt{S=Solver(\ldots)}, \texttt{S.G0\_iw << \ldots} and \texttt{S.solve()}.

The constructor takes the following variables as input:
\begin{description}
    \item[beta] the inverse temperature $\beta$,
    \item[gf\_struct] the Green's function structure,
    \item[Hk] a BlockGf containing the lattice Hamiltonian $H(\kv)$,
    \item[n\_iw] the number of fermionic Matsubara frequencies for $\Delta$ and $G$,
    \item[n\_iw2] the number of fermionic Matsubara frequencies for $G^2$,
    \item[n\_iW] the number of bosonic Matsubara frequencies for $G^2$.
\end{description}
The inverse temperature, \texttt{gf\_struct} and number of frequencies should match those used in the impurity solver. The lattice Hamiltonian $H(\kv)$ is given in momentum space. The momentum mesh of $H(\kv)$ will also be used for all other momentum resolved quantities, e.g., the output $G(\nu,\kv)$ and $\Sigma(\nu,\kv)$.

Prior to running the dual perturbation theory, the following Gf objects should be assigned:
\begin{description}
 \item[Delta] the hybridization function of the impurity model,
 \item[gimp] the impurity single-particle Green's function $g$, as measured by the impurity solver,
 \item[G2\_iw] the two-particle correlation function in the particle-hole AABB notation (in the same convention as used by pomerol2triqs and \cthyb). This is optional, without it no dual self-energy diagrams will be calculated, i.e., the program performs a DMFT calculation.
\end{description}
The following parameters can be given when the dual perturbation theory is run:
\begin{description}
 \item[calculate\_sigma] should the dual self-energy be calculated? If not, skip the three blocks in the middle of Fig.~\ref{fig:flowchart}(b).
 \item[calculate\_sigma1] should the first-order diagram to the dual self-energy be calculated?
 \item[calculate\_sigma2] should the second-order dual self-energy be calculated?
 \item[verbosity] amount of file/textual output to generate. Integer between 0 and 5. This part of the interface is likely to change in future version, with individual booleans controlling aspects of the output.
 \item[delta\_inital] Option for generating an initial guess for DMFT calculations.
\end{description}

\subsection{Impurity solver}

Solving an auxiliary impurity problem is an important part of the dual fermion self-consistency loop. 
This part of the calculation is done by a dedicated application, the impurity solver.
The code has been tested with two impurity solvers available with \triqs, namely \cthyb\ and pomerol. 
Other \triqs-based solvers that expose a similar interface should also work straightforwardly. 
The source code includes tests based on explicit impurity models. These were solved exactly using pomerol, since this allows for reproducible tests. For the tests, the pomerol results are available in the repository, so that the tests can be run without having an impurity solver installed.

\subsection{Explicit expressions for the diagrams}
\label{sec:diagrams}

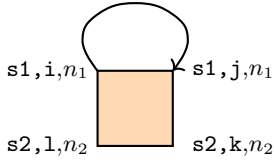
\begin{figure}
\begin{tikzpicture}
 
 \draw[thick,fill=orange!30] (0,0) -- (1,0) -- (1,1) -- (0,1) -- cycle;
 
 \draw[thick,->] plot [smooth,tension=0.8] coordinates {(0,1) (-0.2,1.4) (0,1.75) (0.5,1.9) (1,1.75) (1.2,1.4) (1,1)};
 
 \node[left] at (0,0) {\texttt{s2,l},$n_2$};
 \node[right,xshift=4] at (1,0) {\texttt{s2,k},$n_2$};
 \node[left] at (0,1) {\texttt{s1,i},$n_1$};
 \node[right,xshift=4] at (1,1) {\texttt{s1,j},$n_1$};
 
\end{tikzpicture}
\caption{First-order diagram.}
\label{fig:1st}
\end{figure}

Section 6.2 of Hafermann~\cite{Hafermannphd} contains more details on the derivation of the dual Feynman diagrams. Here, we simply provide explicit expressions for the first two diagrams.

The first-order diagram is given by (see also Fig.~\ref{fig:1st})
\begin{align}
 \tilde{\Sigma}^\texttt{s2}_{\texttt{l,k}}&(n_2,\Rv) =\\ &\frac{-1}{\beta} \delta_{\Rv} \sum_{\texttt{i,j}} \sum_{n_1,n_2} \sum_\texttt{s1} \gamma^{\texttt{s1},\texttt{s2}}_{\texttt{i,j,k,l}}(0,n_1,n_2)\tilde{G}^{\texttt{s1}}_{\texttt{i,j}}(n_1,\Rv). \notag
\end{align}
Here, the arguments are used as follows: the upper index $^\texttt{s1}$ is the block index, the lower indices $_\texttt{i,j}$ are orbital indices, $n_1,n_2$ are fermionic Matsubara frequencies and $\Rv$ is a vector in real space. 

\begin{figure}
\begin{tikzpicture}
 
  \draw[thick,fill=orange!30] (0,0) -- (1,0) -- (1,1) -- (0,1) -- cycle;

   \draw[thick,fill=orange!30] (4.5,0) -- (5.5,0) -- (5.5,1) -- (4.5,1) -- cycle;
   
   \draw[thick,->] (1,0) -- node[below]{\texttt{s2,Ak$\rightarrow$Bl},$n_2+w$,$+\Rv$} (4.5,0) ;

   \draw[thick,<-] (1,1) -- node[below]{\texttt{s1,Bi$\rightarrow$Aj},$n_1+w$,$-\Rv$} (4.5,1) ;
   
   \draw[thick,->] plot [smooth,tension=0.3] coordinates {(0,1) (-0.2,1.4) (0,1.75) (2.75,1.9) (5.5,1.75) (5.7,1.4) (5.5,1)};
   
   \node[left] at (0,0) {\texttt{s2,Al},$n_2$};
   \node[right] at (5.5,0) {\texttt{s2,Bk},$n_2$};

   \node[above] at (2.75,1.9) {\texttt{s1,Ai$\rightarrow$Bj},$n_1$,$+\Rv$};
   
\end{tikzpicture}
\\
\vspace{0.5cm}
\begin{tikzpicture}
 
  \draw[thick,fill=orange!30] (0,0) -- (1,0) -- (1,1) -- (0,1) -- cycle;

   \draw[thick,fill=orange!30] (4.5,0) -- (5.5,0) -- (5.5,1) -- (4.5,1) -- cycle;
   
   \draw[thick,->] (1,0) -- node[below]{\texttt{s2,Ai$\rightarrow$Bj},$n_1$,$+\Rv$} (4.5,0) ;

   \draw[thick,<-] (1,1) -- node[below]{\texttt{s1,Bi$\rightarrow$Aj},$n_1+w$,$-\Rv$} (4.5,1) ;
   
   \draw[thick,->] plot [smooth,tension=0.3] coordinates {(0,1) (-0.2,1.4) (0,1.75) (2.75,1.9) (5.5,1.75) (5.7,1.4) (5.5,1)};
   
   \node[left] at (0,0) {\texttt{s2,Al},$n_2$};
   \node[right] at (5.5,0) {\texttt{s2,Bk},$n_2$};

   \node[above] at (2.75,1.9) {\texttt{s1,Ak$\rightarrow$Bl},$n_2+w$,$+\Rv$};
   
\end{tikzpicture}
\caption{Second-order diagram with several possible assignments of indices.}
\label{fig:2nd}
\end{figure}
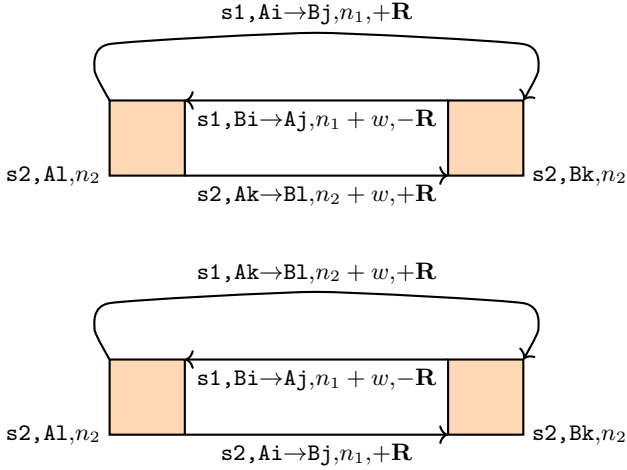

At second-order, there are two ways to assign the labels, as shown in Fig.~\ref{fig:2nd}. The resulting self-energy is
\begin{widetext}
\begin{align}
 \tilde{\Sigma}&^\texttt{s2}_{\texttt{Al,Bk}}(n_2,\Rv) \\
 =& \frac{-1}{2\beta^2} \sum_{\texttt{Ai,Aj,Ak,Bi,Bj,Bl}} \sum_{w,n_1,n_2} \sum_\texttt{s1}
 \gamma^{\texttt{s1},\texttt{s2}}_{\texttt{Ai,Aj,Ak,Al}}(w,n_1,n_2)
 \gamma^{\texttt{s1},\texttt{s2}}_{\texttt{Bi,Bj,Bk,Bl}}(-w,n_1+iw,n_2+iw) \notag \\
 &\times
 \tilde{G}^{\texttt{s1}}_{\texttt{Ai,Bj}}(n_1,\Rv) 
 \tilde{G}^{\texttt{s1}}_{\texttt{Bi,Aj}}(n_1+w,-\Rv)  
 \tilde{G}^{\texttt{s2}}_{\texttt{Ak,Bl}}(n_2+iw,\Rv)  \notag \\
 &+
 \frac{-1}{2\beta^2} \sum_{\texttt{Ai,Aj,Ak,Bi,Bj,Bl}} \sum_{n_1,n_2,w} \sum_\texttt{s1}
 \gamma^{\texttt{s1},\texttt{s2}}_{\texttt{Ak,Aj,Ai,Al}}(n_1-n_2,n_2+w,n_2)
 \gamma^{\texttt{s1},\texttt{s2}}_{\texttt{Bi,Bl,Bk,Bj}}(n_2-n_1,n_1+w,n_1) \notag \\
 &\times (1-\delta_{\texttt{s1,s2}})
 \tilde{G}^{\texttt{s1}}_{\texttt{Ak,Bl}}(n_2+w,\Rv) 
 \tilde{G}^{\texttt{s1}}_{\texttt{Bi,Aj}}(n_1+w,-\Rv)  
 \tilde{G}^{\texttt{s2}}_{\texttt{Ai,Bj}}(n_1,\Rv). \notag  
\end{align}

\end{widetext}

\subsection{Parallelization}

MPI parallelization is implemented by splitting up the internal frequency sum in the evaluation of the self-energy diagrams. Parallelization automatically occurs over the entire frequency mesh (one bosonic and two fermionic Matsubara frequencies). 

No attempt is made to limit the memory use per process in the parallelization. In situations where the total amount of memory is an important constraint (many orbitals, many frequencies, dense momentum grid), it can be useful to limit the number of parallel processes per compute node. 

\section{SrVO3}
\label{sec:srvo3}

\begin{figure*}
 \includegraphics{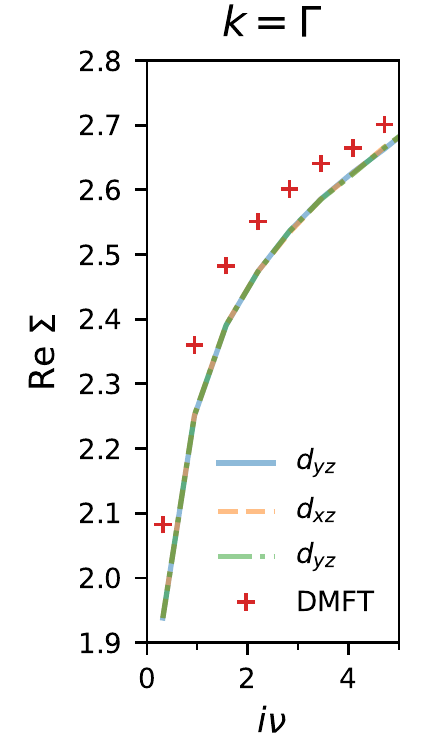}%
 \includegraphics{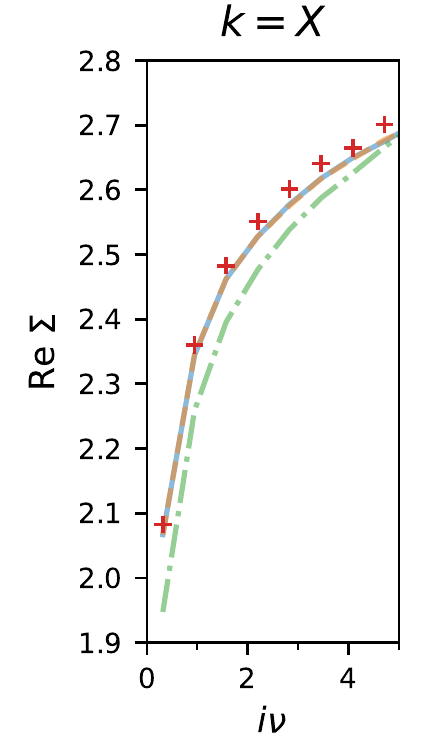}%
 \includegraphics{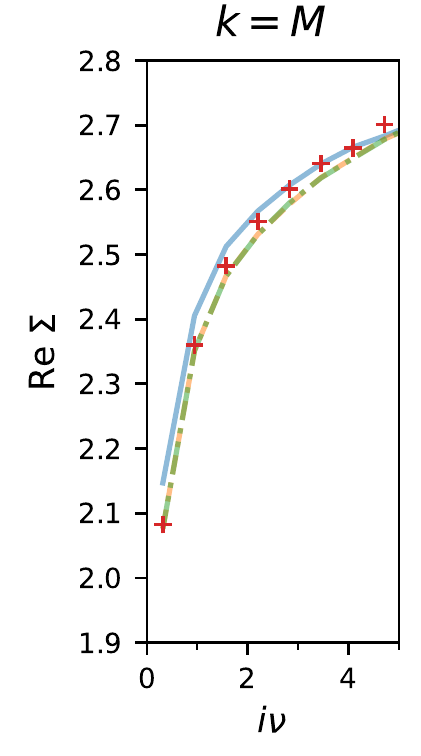}%
 \includegraphics{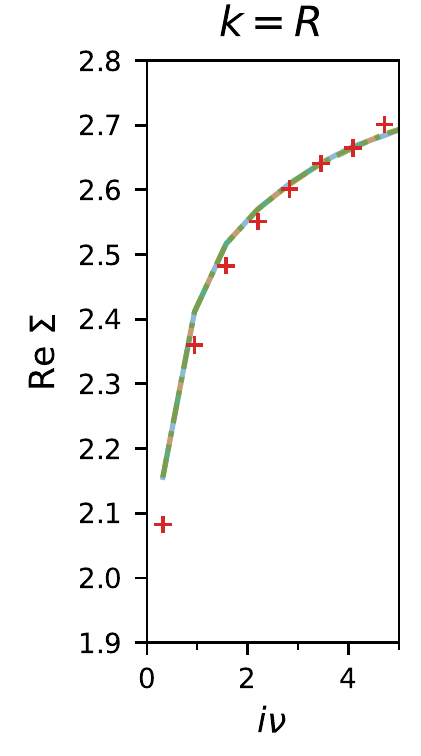}\\
 \includegraphics{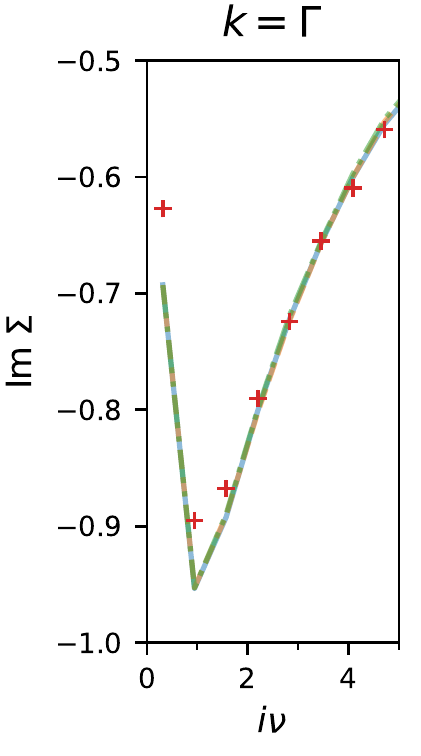}%
 \includegraphics{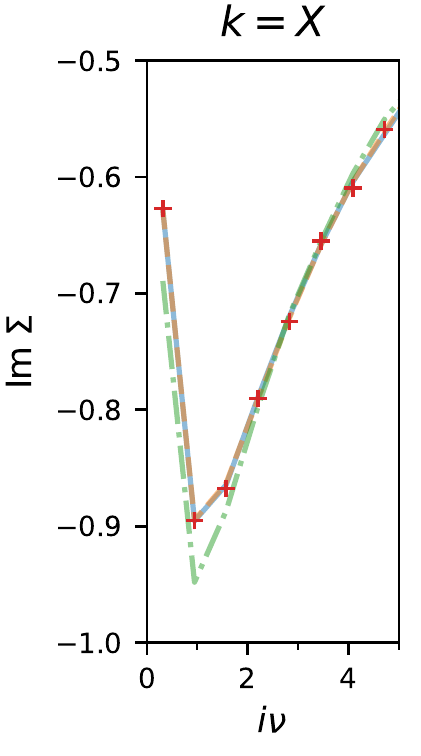}%
 \includegraphics{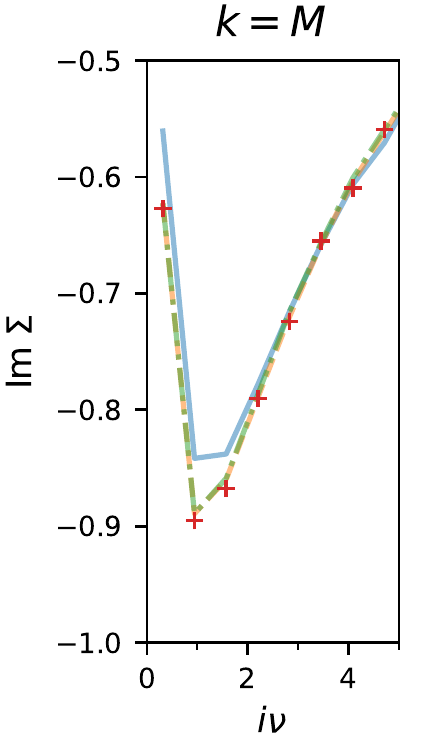}%
 \includegraphics{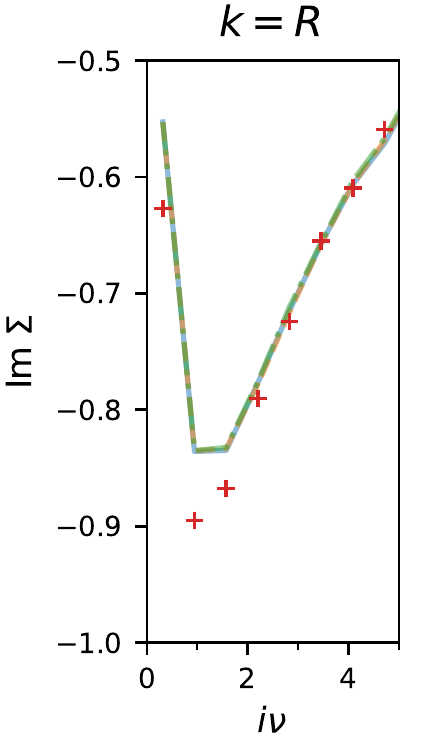}
 \caption{Self-energy for SrVO$_3$ at the high-symmetry momenta $\Gamma=(0,0,0)$, $X=(\pi,0,0)$, $M=(\pi,\pi,0)$ and $R=(\pi,\pi,\pi)$.}
 \label{fig:srvo3:sigma_k}
\end{figure*}

\begin{figure}
 \includegraphics{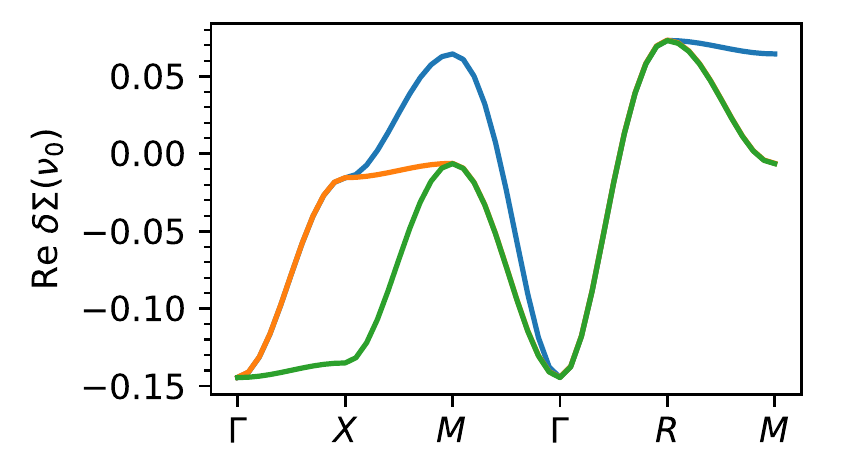}\\
 \includegraphics{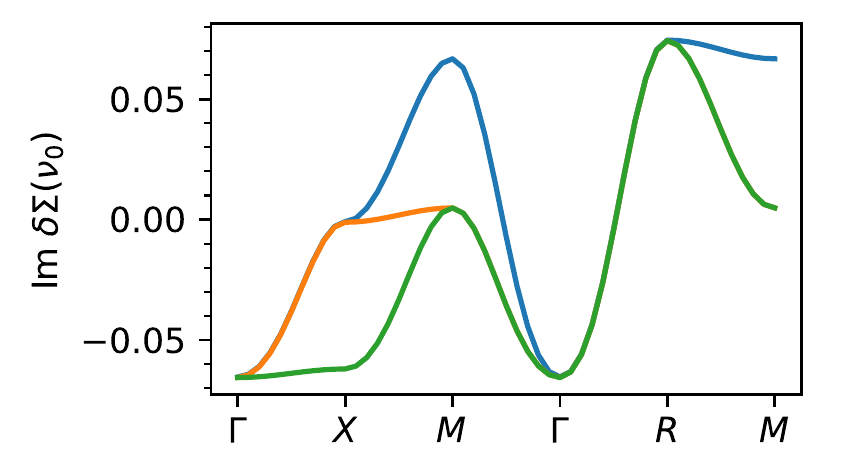}%
 \caption{Change in self-energy $\delta \Sigma = \Sigma-\Sigma^\text{DMFT}$ at the lowest Matsubara frequency for SrVO$_3$.}
 \label{fig:srvo3:sigma_path}
\end{figure}

To illustrate the technique, we investigate SrVO$_3$, which is a popular material in the study of electronic correlations. The spatial correlations beyond DMFT have previously been investigated using the dynamical vertex approximation~\cite{Galler17,Galler19,Kaufmann20}. To facilitate the comparison, we use the non-interacting Hamiltonian $H(\kv)$ of Ref.~\onlinecite{Galler19}, with $20\times 20\times 20$ k-points, the same Kanamori interaction $U=5$ eV, $V=3.5$ eV, $J=0.75$ eV, and the same inverse temperature $\beta = 10$ eV.

The Green's function consists of two spin blocks with three t2g orbitals each. The t2g orbitals are related by rotational symmetry and their \emph{local} Green's functions are thereby equivalent. The calculations are done with the full $3\times 3$ matrix structure, but \emph{a posteriori} the self-energy turns out to be almost completely orbital diagonal and only the diagonal components are shown in the figures. 

The impurity model is solved using \cthyb. A moderately small frequency box for the vertex, $15\times 16 \times 16$~\footnote{I.e., with \cthyb\ parameters \texttt{measure\_G2\_n\_bosonic: 8,
        measure\_G2\_n\_fermionic: 8}}, is sufficient to ensure convergence of the self-energy for the frequencies shown in the figures. A larger number of frequencies is needed to determine (especially) $\Re \Sigma(i\nu)$ at large $i\nu$ accurately.

Figures~\ref{fig:srvo3:sigma_k} and \ref{fig:srvo3:sigma_path} show that the self-energy has a non-local part. As a function of momentum, the real part of the self-energy varies by approximately 0.2 eV,  the change in the imaginary part is somewhat smaller. These magnitudes are consistent with the previous D$\Gamma$A studies~\cite{Galler19}. 

In our results, the change in self-energy $\delta \Sigma = \Sigma(\kv,\nu) - \Sigma^\text{DMFT}(\nu)$ is distributed rather evenly around zero. In other words, the dual fermion corrections do not (substantially) change the local part of the self-energy. This is consistent with Fig.~5 of Ref.~\cite{Galler19}.

From the dual fermion theory~\cite{Rubtsov08}, a small change in local self-energies is expected, since DMFT is assumed to provide an accurate description of the local correlations. In the second order of dual perturbation theory, the local dual self-energy is proportional to the local dual Green's function, which is zero if the DMFT hybridization is taken~\cite{Rubtsov09}. Thus, local self-energy corrections only appear through the transformation from the dual to the lattice self-energy.

Figure~\ref{fig:srvo3:sigma_path} shows that the combined orbital-momentum structure of the self-energy is reminiscent of the original band structure and can be interpreted as a band-widening~\cite{Miyake13,Tomczak14,Galler17} effect.

The non-local corrections are largely restricted to the lowest Matsubara frequencies. This is especially the case for the imaginary part. In the dual theory, this phenomenon stems from frequency structure $\tilde{G} \sim G_{\kv\nu}-g_\nu$ of the dual Green's function~\cite{Hafermannphd}.
        
The total runtime of the dual perturbation theory for a calculation with \texttt{measure\_G2\_n\_bosonic=8} and \texttt{measure\_G2\_n\_fermionic}=8 was just under 9 hours per process distributed over 50 processes. Due to the large number of k-points, memory is a main bottleneck for this calculation, requiring approximately 25 GB per process. Solving the impurity model using \cthyb\ was done for 16 hours on 400 processes.
        
\section{Testing against exactly solvable small systems: the Hubbard dimer in the two exact dual fermion limits}
\label{sec:clustertesting}

To illustrate and test the code, it is useful to study small clusters.
The idea is that these are small enough to be solved in ED, so that the exact self-energy can be obtained at any interaction strength. 
This exact result can then be compared with the result of \DF\ starting from a smaller system. The \DF\ perturbation theory requires the two-particle vertex of the small system coupled to a bath, easily obtained via an ED solver such as pomerol. These tests are sufficiently simple to run quickly in a normal desktop environment and example scripts are provided along with this paper.

The test cases need to be chosen appropriately: Second-order dual perturbation theory is generally not exact, so one cannot always expect complete agreement with the reference data. 
However, the dual perturbation theory is rapidly convergent -- so that second order suffices -- in two opposite limits: weak interaction $U$ or weak hopping $t$ (between impurities). To obtain the exact result up to second order in $U$ or $t$, it is necessary to use an appropriate bath. With that in place, there should be quantitative agreement between dual perturbation theory and the reference model and this acts as a stringent test on the implementation.

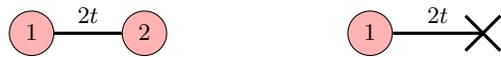
\begin{figure}
\begin{tikzpicture}[
site/.style={circle,draw=black,text height=1.5ex,text depth=0.25ex},
s1/.style={fill=red!30},
s2/.style={fill=blue!30},
bath/.style={inner sep=6,very thick,draw,circle,cross out},
t0/.style={very thick},
t1/.style={thick,densely dotted},
]
 
\begin{scope}[scale=1.5]

\node[site,s1] (1) at (0,0) {1};
\node[site,s1] (2) at (1,0) {2};

\draw[t0] (1) -- node[above] {$2t$} (2) ;

\end{scope}

\begin{scope}[scale=1.5,shift={(3,0)}]

\node[site,s1] (1) at (0,0) {1};
\node[bath] (2) at (1,0) {};

\draw[t0] (1) -- node[above] {$2t$} (2.center) ;

\end{scope}
\end{tikzpicture}
\caption{Hubbard dimer (left) and auxiliary impurity model (right).}
\label{fig:dimer}
\end{figure}

\begin{figure*}
\includegraphics[width=0.25\textwidth]{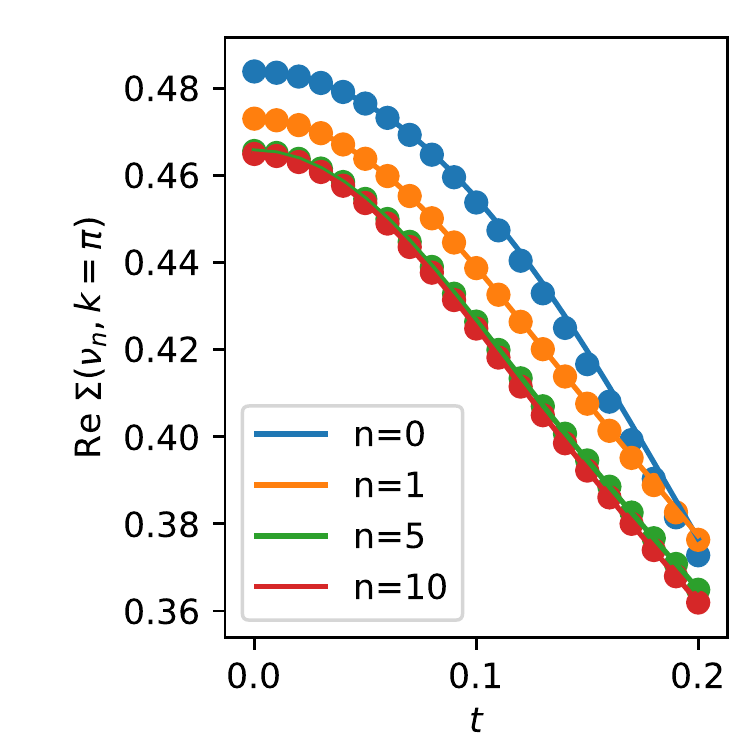}%
\includegraphics[width=0.25\textwidth]{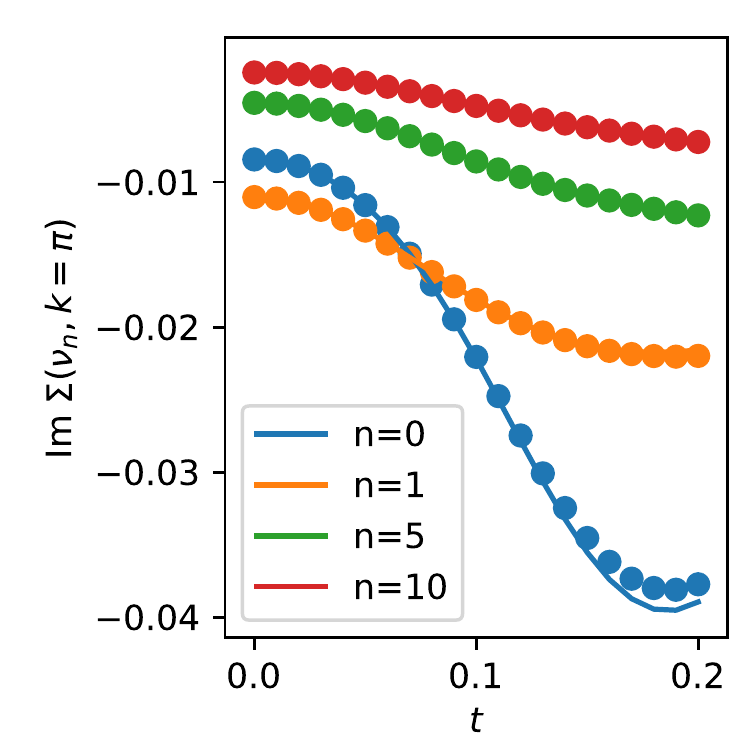}%
\includegraphics[width=0.25\textwidth]{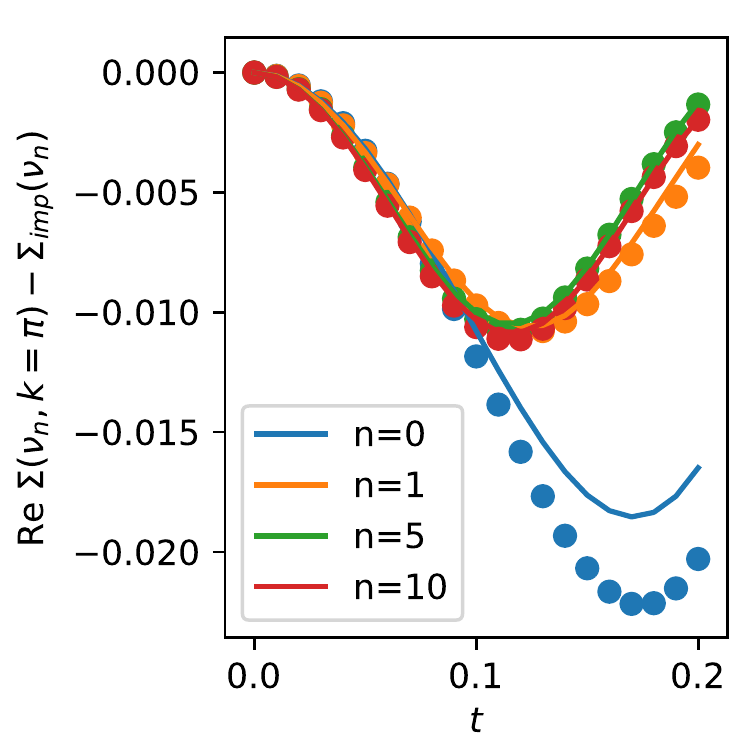}%
\includegraphics[width=0.25\textwidth]{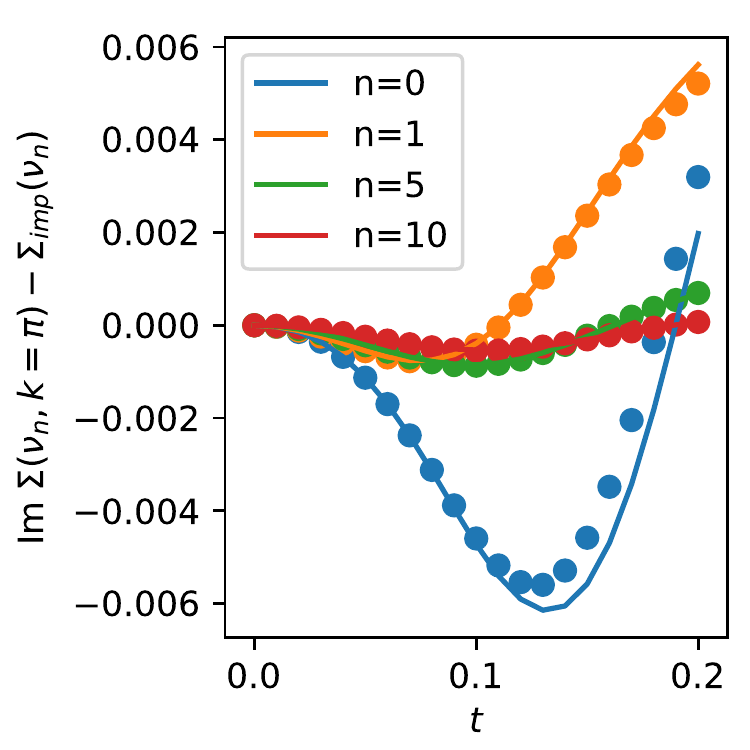}%
\caption{Scaling of the self-energy at small $t$. $U=0.5$, $\mu=0.75$ (out of half-filling), $\beta=10$. Second-order pertubation theory (symbols) and the exact result (lines) agree to order $t^2$. At higher order, deviations start to appear, especially in the real part at low Matsubara frequency $\nu_n$.}
\label{fig:2site:smallt}
\end{figure*}

We start with the half-filled Hubbard dimer, shown in Fig.~\ref{fig:dimer}, the smallest system that features non-local correlations. The Hamiltonian in real space is
\begin{align}
 H = -2t \sum_{\sigma=\up,\dn} \left[ c^\dagger_{1\sigma} c^{\phantom{\dagger}}_{2,\sigma} + c^\dagger_{2\sigma} c^{\phantom{\dagger}}_{1,\sigma} \right] +U\left[n_{1\up}n_{1\dn}+n_{2\up}n_{2\dn}\right].
\end{align}
The single-particle hopping can also be written in momentum space, as
\begin{align}
 H = \sum_{\sigma=\up,\dn} \sum_{k_x=0,\pi} -2 t \cos(k_x) c^\dagger_{\sigma,k_x} c^{\phantom{\dagger}}_{\sigma,k_x}+U\left[n_{1\up}n_{1\dn}+n_{2\up}n_{2\dn}\right].
\end{align}
The right-hand side of Fig.~\ref{fig:dimer} shows an auxiliary impurity model with a single interacting site (red circle) and a bath consisting of a single non-interacting site (black cross). The corresponding Hamiltonian is
\begin{align}
 H = 
 -2t \sum_{\sigma=\up,\dn} \left[ c^\dagger_{1\sigma} c^{\phantom{\dagger}}_{b,\sigma} + c^\dagger_{b\sigma} c^{\phantom{\dagger}}_{1,\sigma} \right]
 +Un_{1\up}n_{1\dn}. 
\end{align}
The corresponding hybridization function is 
$\Delta(i\nu_n) = \frac{(2t_1)^2}{i\nu_n}$. From this impurity model, we obtain the impurity Green's function $\av{c_1^{\phantom{\dagger}}c_1^\dagger}$. Note that the construction of the impurity model guarantees that the auxiliary impurity Green's function is equal to the exact local Green's function at $U=0$.

\begin{figure*}
\includegraphics[width=0.25\textwidth]{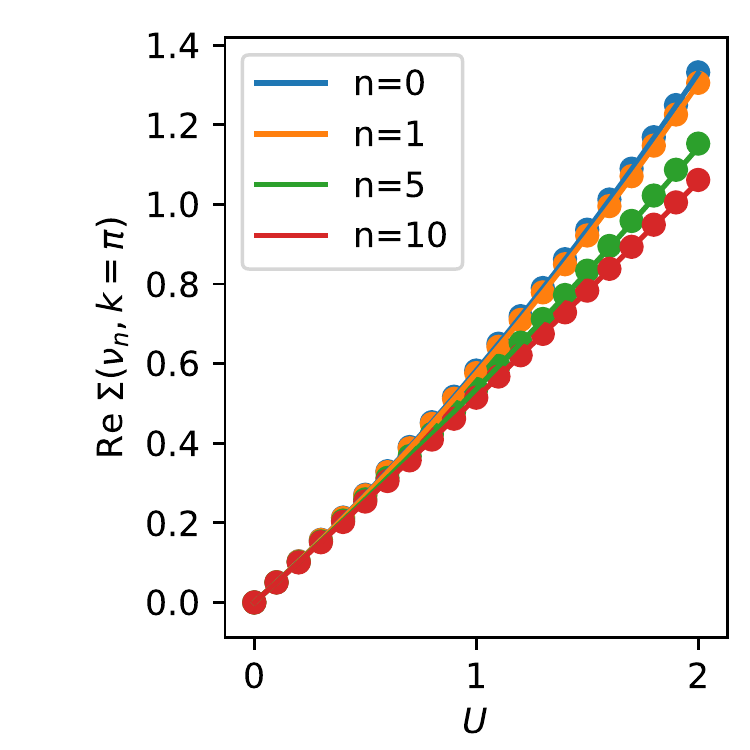}%
\includegraphics[width=0.25\textwidth]{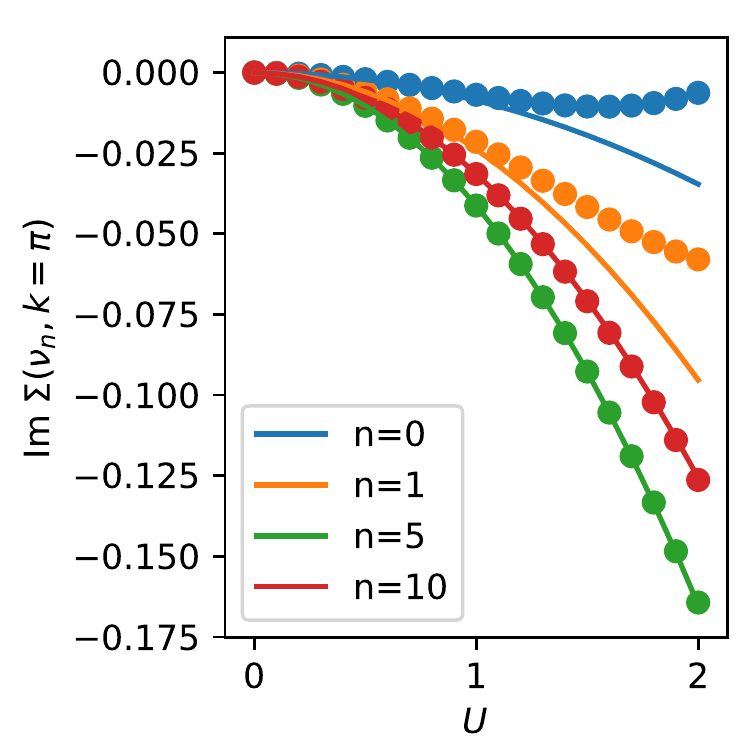}%
\includegraphics[width=0.25\textwidth]{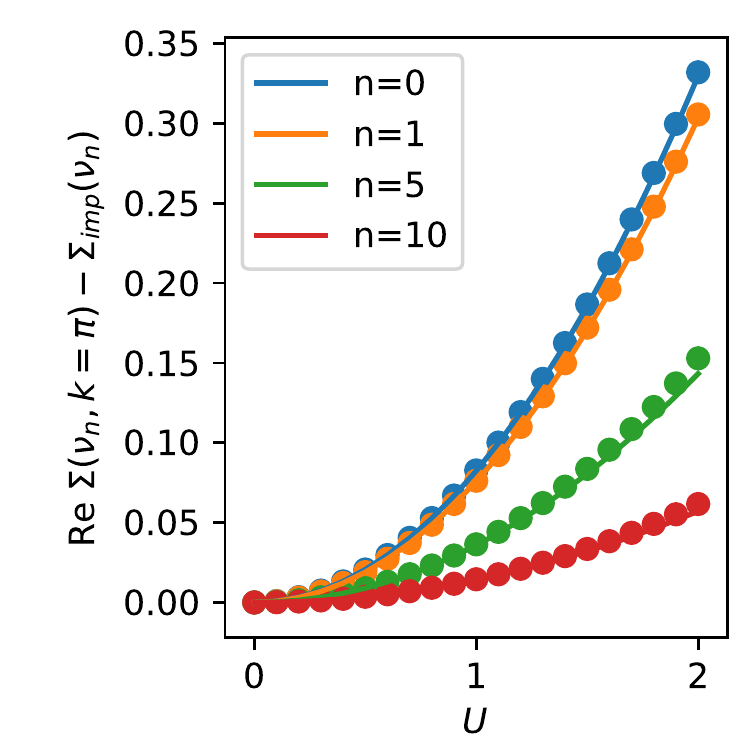}%
\includegraphics[width=0.25\textwidth]{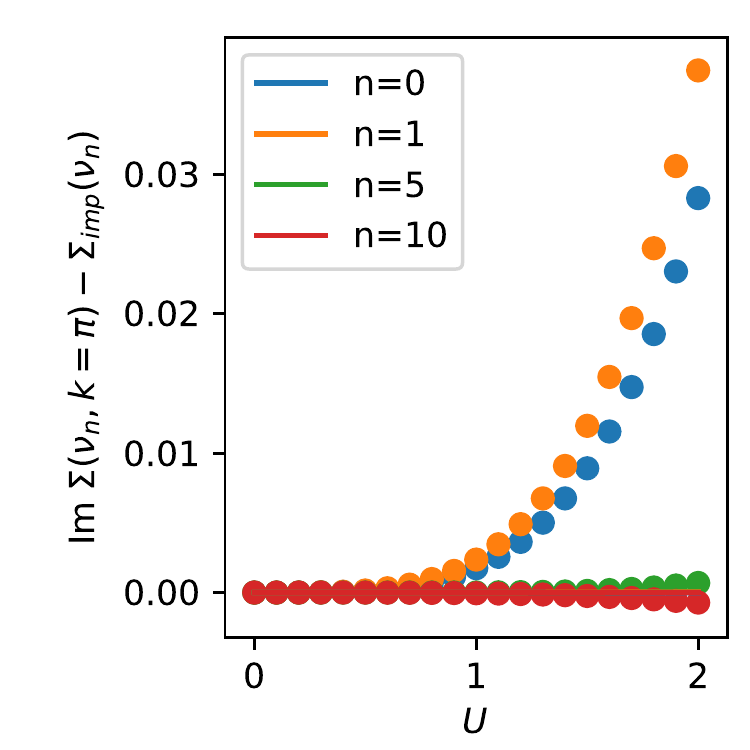}%
\caption{Scaling of the self-energy at small $U$. $t=0.5$, $\mu=U/2$ (half-filling), $\beta=10$. Second-order pertubation theory (symbols) and the exact result (lines) agree to order $U^2$ at small $U$. At larger $U$, deviations start to appear, especially in the imaginary part at low Matsubara frequency $\nu_n$.}
\label{fig:2site:smallU}
\end{figure*}

This forms the starting point for the proof that the dual fermion perturbation theory based on this impurity model is exact to order $U^2$. The chosen auxiliary model guarantees that $\tilde{G}(\Rv=0)=0$ at $U=0$, in other words, $\tilde{G}(\Rv=0) \propto U$ (at half-filling, the odd term vanishes due to particle-hole symmetry and $\tilde{G}(\Rv=0) \propto U^2$). The $n$-fermion dual interaction (with $2n$ legs) scales as $U^{n-1}$, so the highest order that needs to be considered is the three-particle interaction. However, the only diagram with a single three-fermion interaction also contains two local dual Green's functions, raising the order in $U$. As a result, only diagrams with 2-fermion interactions (and only up to 2 of those) can contribute at order $U^2$. These are exactly the diagrams included in this code.

In the opposite limit of small $t/U$, we expand the self-energy in $t$ up to $t^3$. At $t=0$, the exact Green's function is completely local and equal to the impurity Green's function. We have $\tilde{G}(\Rv)=0$ at all $\Rv$. In other words, at small $t$, every Green's function contributes at least one power of $t$, limiting the number of allowed diagrams. Expanding in terms of the number of hopping events $t$ shows that the local part $\tilde{G}(\Rv=0)\propto t^2$ (it requires one hop forward and one hop back). Thus, the lowest-order diagram with a three-fermion vertex, which has two local internal lines, is of order $t^4$. Higher-order diagrams with multi-fermion vertices have at least 4 internal lines and are thus also of order $t^4$. This again leaves the first and second order diagrams as the only diagrams of order $t^3$.

The scaling of the self-energy for small $t$, using $U=0.5$, $\mu=0.75$ and $\beta=10$ is shown in Fig.~\ref{fig:2site:smallt}. Both the real and imaginary part of the self-energy scale quadratically in $t$. The imaginary part decays quickly as a function of the Matsubara frequency $\nu_n$ (shown in various colors), whereas the real part is somewhat independent of $\nu_n$. The quality of the second-order dual approximation becomes clearer by subtracting the impurity self-energy, i.e., by looking at $\Sigma-\Sigma_\text{imp}$. The dual solution follows the exact result at small $t$. At larger $t$ (note: $U=0.5$ sets the energy scale), the dual solution deviates from the exact result, especially at low frequency and in the real part.

For the small $U$ scaling, Fig.~\ref{fig:2site:smallU}, we consider the half-filled dimer ($\mu=U/2$ changes along with the interaction) at $t=0.5$, $\beta=10$. We again find the correct quadratic scaling in $\Im \Sigma$ and linear (Hartree) scaling in $\Re \Sigma$. Substantial deviations in $\Im \Sigma$ start to appear at $U\approx 2t$. The second-order dual perturbation theory suggests that $\Sigma$ is acquiring nonlocal contributions at this point, but the exact self-energy is still almost entirely local. Apparently, some cancellation of higher-order self-energy corrections occurs in the exact system that is not captured by this calculation. Possible origins are higher-order diagrams, higher-order vertices, or the fact that the present construction of the reference model is only optimal in the limit of small $U$.

The dimer is an artificial example that is potentially far away from the mean-field limit. In the regimes studied here, the overall magnitude of the non-local corrections is small compared to the energy scales of the problem. However, the use of ED gives us the ability to verify exact properties of the theory and the implementation.

\section{A 4 site cluster}

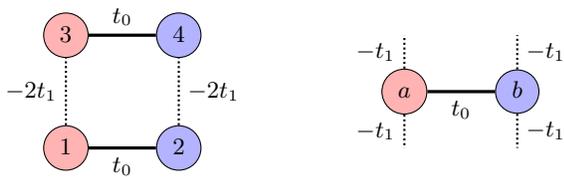
\begin{figure}
\begin{tikzpicture}[
site/.style={circle,draw=black,text height=1.5ex,text depth=0.25ex},
s1/.style={fill=red!30},
s2/.style={fill=blue!30},
t0/.style={very thick},
t1/.style={thick,densely dotted},
]
 
\begin{scope}[scale=1.5]

 \node[site,s1] (1) at (0,0) {1} ;
 \node[site,s2] (2) at (1,0) {2} ;
 \node[site,s1] (3) at (0,1) {3} ;
 \node[site,s2] (4) at (1,1) {4} ;
 
 \draw[t0] (1) -- node[below] {$t_0$} (2) ;
 \draw[t0] (3) -- node[above] {$t_0$} (4) ;
 \draw[t1] (1) -- node[left] {$-2t_1$} (3);
 \draw[t1] (2) -- node[right] {$-2t_1$} (4);
\end{scope}

\begin{scope}[scale=1.5,shift={(3,0.5)}]
 \node[site,s1] (1) at (0,0) {$a$} ;
 \node[site,s2] (2) at (1,0) {$b$} ;
 \draw[t0] (1) -- node[below] {$t_0$} (2) ; 
 \draw[t1] (1) -- node[left] {$-t_1$} ++(0,0.5); 
 \draw[t1] (1) -- node[left] {$-t_1$} ++(0,-0.5); 
 \draw[t1] (2) -- node[right] {$-t_1$} ++(0,0.5); 
 \draw[t1] (2) -- node[right] {$-t_1$} ++(0,-0.5); 
\end{scope}

\end{tikzpicture}
\caption{A cluster consisting of four sites (left) and a one-dimensional periodic lattice of dimers (right) are equivalent when $N_\text{dimers}=2$.}
\label{fig:4site}
\end{figure}

To further illustrate and test the code, it is useful to consider a small cluster of four sites that is decomposed into two dimers. Unlike the previous example, this allows us to investigate the multi-orbital structure of the code in a minimal model.

An example of a four site cluster is shown in Fig.~\ref{fig:4site} (left). Every site contains one orbital. There is a Hubbard interaction $U_a$ on sites 1 and 3 (red) and another Hubbard interaction $U_b$ on sites 2 and 4 (blue). Hopping occurs both between sites of the same color ($t_1$) and sites of different color ($t_0$). The Hamiltonian is
\begin{align}
 H =& -2 t_1 \sum_{\sigma=\up,\dn} \left[ 
 c^\dagger_{1\sigma} c^{\phantom{\dagger}}_{3\sigma} +
 c^\dagger_{3\sigma} c^{\phantom{\dagger}}_{1\sigma} +
 c^\dagger_{2\sigma} c^{\phantom{\dagger}}_{4\sigma} +
 c^\dagger_{4\sigma} c^{\phantom{\dagger}}_{2\sigma}  
 \right]  
 \notag \\
 &+t_0 \sum_{\sigma=\up,\dn} \left[ 
 c^\dagger_{1\sigma} c^{\phantom{\dagger}}_{2\sigma} +
 c^\dagger_{2\sigma} c^{\phantom{\dagger}}_{1\sigma} +
 c^\dagger_{3\sigma} c^{\phantom{\dagger}}_{4\sigma} +
 c^\dagger_{4\sigma} c^{\phantom{\dagger}}_{3\sigma}  
 \right] \notag \\
 &+
 U_a \left[ n_{1\up} n_{1\dn} + n_{3\up} n_{3\dn} \right] \notag \\
 &+
 U_b \left[ n_{2\up} n_{2\dn} + n_{4\up} n_{4\dn} \right]
\end{align}

The same system can be seen as a one-dimensional periodic lattice with a dimer (two-orbital) unit cell, Fig.~\ref{fig:4site} (right), with $N_\text{dimer}=2$ dimers in total. The multi-orbital \DF\ method can be applied straightforwardly in this formulation. The 'atomic' Hamiltonian is
\begin{align}
 H_\text{at} =& +t_0 \sum_{\sigma=\up,\dn} \left[ 
 c^\dagger_{a\sigma} c^{\phantom{\dagger}}_{a\sigma} +
 c^\dagger_{b\sigma} c^{\phantom{\dagger}}_{b\sigma}  
 \right] \notag \\
 &+
 U_a n_{a\up} n_{a\dn}+
 U_b n_{b\up} n_{b\dn}
\end{align}
The original Hamiltonian is written as a lattice Hamiltonian as a sum over unit cells (``atoms''), indexed by Roman capitals, and orbital indices, indexed by Greek lowercase letters.
\begin{align}
 H =& \sum_{J\in\{0,1\}} H_{\text{at,}J} \notag \\
 &+\sum_{\sigma=\up,\dn}\sum_{I,J \in \{0,1\} }\sum_{\alpha,\beta \in \{a,b\} }  \hat{t}_{IJ,\alpha\beta}  c^\dagger_{I\alpha\sigma} c^{\phantom{\dagger}}_{J\beta\sigma},
\end{align}
where $\hat{t}$ is the interdimer dispersion
\begin{align}
 \hat{t}_{IJ,\alpha\beta} = \begin{cases}
                             -2t_1 \text{, if }I\neq J \text{ and } \alpha\neq \beta \\
                             0
                            \end{cases}.
\end{align}
Performing a Fourier transform from $I$, $J$ to the one-dimensional momentum $k$ and using a matrix in orbital space,
\begin{align}
 \hat{t}(k) =& \begin{pmatrix}
              -2t_1 \cos(k)  & 0 \\
              0 & -2t_1 \cos(k) 
              \end{pmatrix} \\
 H =& \sum_{J\in\{0,1\}} H_{\text{at,}J} +\sum_{\sigma=\up,\dn}\sum_{k \in \{0,\pi\} }\sum_{\alpha,\beta \in \{a,b\} } 
 c^\dagger_{k\alpha\sigma} 
 \,\,
 \hat{t}(k)_{\alpha\beta}  
 \,\,
 c^{\phantom{\dagger}}_{k\beta\sigma}.
\end{align}
Note that we consider an isolated dimer as the impurity model, i.e., without any bath states.

\begin{figure*}
\includegraphics{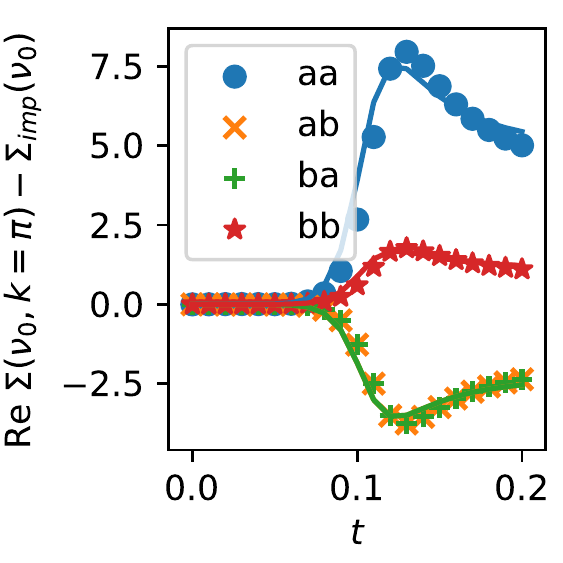}%
\includegraphics{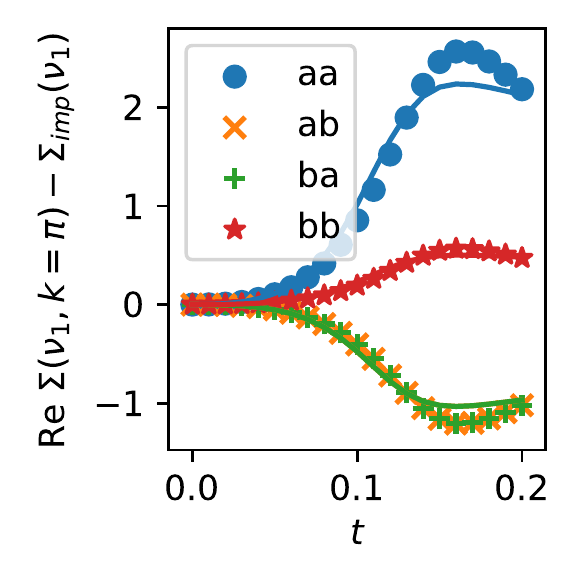}%
\includegraphics{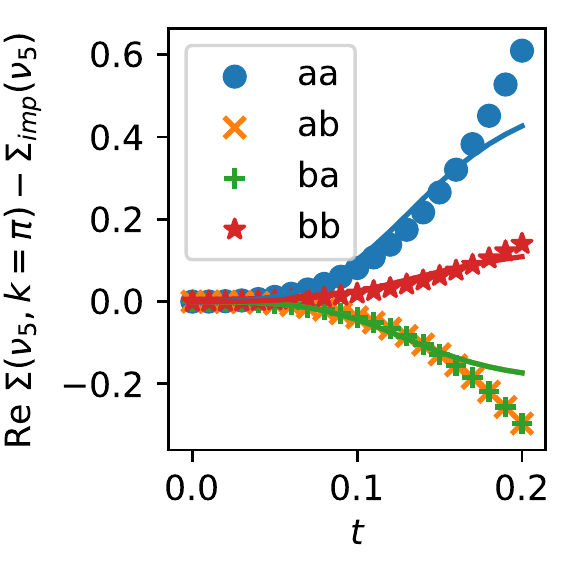}\\
\includegraphics{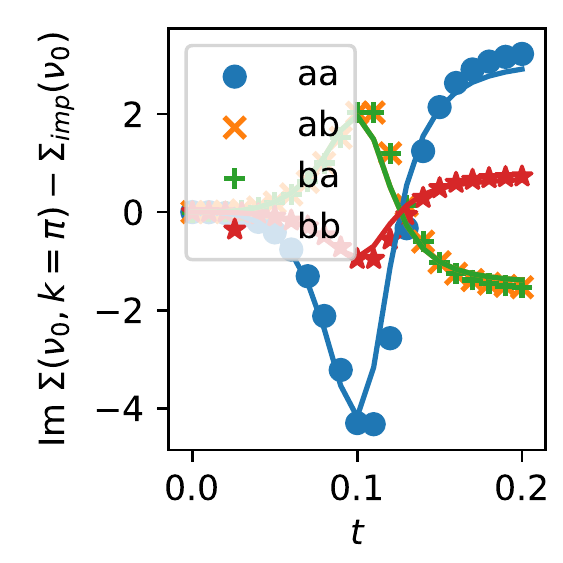}%
\includegraphics{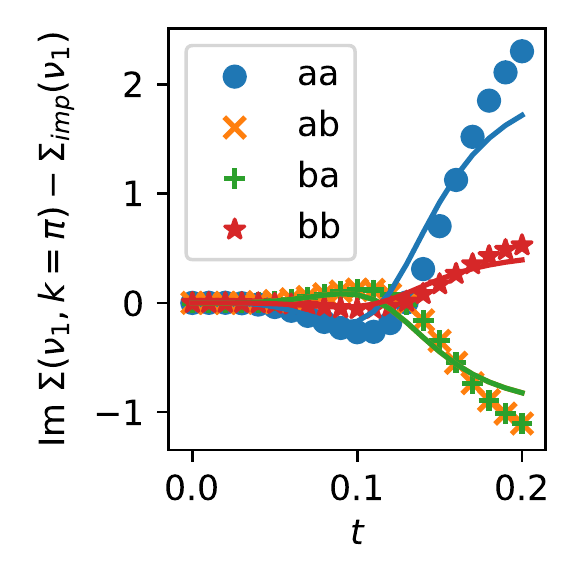}%
\includegraphics{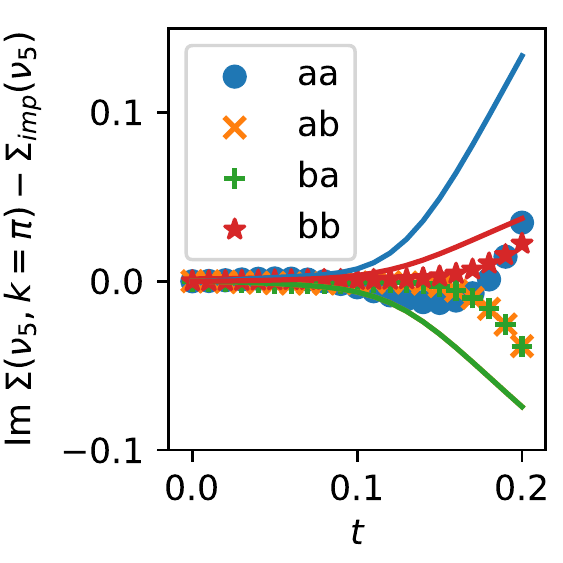}
\caption{Correction $\Sigma(k)-\Sigma(\text{dimer})$ to the dimer self-energy for $U_a=10$, $U_b=3$, $t_0=4$, $\beta=20$, as a function of $t_1$. The symbols show the dual fermion result, the lines are the exact result obtained by diagonalizing the full cluster. From left to right, the frequencies $\nu_0$, $\nu_1$ and $\nu_5$ are shown. The different symbols/colors correspond to orbital components of the self-energy in the $2\times 2$ matrix notation.}
\label{fig:cluster1}
\end{figure*}

\subsubsection{Definition of the self-energy from ED}

The goal is to compare the self-energy from the ED solution with the second-order dual fermion approximation. To do this, we transform the ED Green's function to the momentum-orbital basis,
\begin{align}
 &G_{\alpha\beta}(k) \leftrightarrow G_{ij} \pm G_{i,j+2} \\
 &\text{ with }
 \begin{cases}
  i=1 \text{ if } \alpha = a \\
  i=2 \text{ if } \alpha = b
 \end{cases}
 \begin{cases}
  j=1 \text{ if } \beta = a \\
  j=2 \text{ if } \beta = b
 \end{cases}
\end{align}
where the $\pm$ is $+$ for $k=0$ and $-$ for $k=\pi$. The non-interacting Green's function is
\begin{align}
 \left[ G^0(k) \right]^{-1}_{\alpha\beta} =& (i\nu+\mu) \hat{1} - \epsilon_{\alpha\beta}(k) \\
 &=
 \begin{pmatrix}
      i\nu+\mu+2t_1 \cos(k)  & -t_0 \\
      -t_0 & i\nu+\mu+2t_1 \cos(k)  
 \end{pmatrix}, \notag
\end{align}
where $\epsilon$ is the full dispersion,
\begin{align}
 \epsilon_{\alpha\beta}(k) =  
      \begin{pmatrix}
      -2t_1 \cos(k)  & +t_0 \\
      +t_0 & -2t_1 \cos(k) 
      \end{pmatrix}.
\end{align}
The self-energy is obtained as the difference of inverse Green's functions according to Dyson's equation
\begin{align}
 \Sigma(k)_{\alpha\beta} = \left[ G^0(k) \right]^{-1}_{\alpha\beta} - \left[ G(k) \right]^{-1}_{\alpha\beta}.
\end{align}

\subsubsection{Intra-dimer hopping in the CTHYB+DF2 program}

The intra-dimer hopping $t_0$ needs to be included in the computation.
The \DF\ program requires $\hat{t}$ without local part, so it cannot be included there and this means it should also not be included in $\Delta$.
Instead, we include all intra-dimer quadratic terms into $g_0$, 
\begin{align}
 \left[ g_0(i\nu_n)\right]^{-1} = \left(i \nu_n + \mu\right) \hat{1} -       
  \begin{pmatrix}
  0  & +t_0 \\
  +t_0 & 0 
  \end{pmatrix}
  - \hat{\Delta}(i\nu_n).
\end{align}

\subsubsection{Results}

The results of this test are shown in Figs.~\ref{fig:cluster1}. The self-energy in the lattice-of-dimers representation is $\Sigma(k)_{\alpha\beta}$ with $\alpha,\beta \in \{a,b\}$ and $k\in\{0,\pi\}$. The figures show the correction to the dimer self-energy, i.e., $\Sigma(k) - \Sigma(\text{dimer})$ for both dual perturbation theory (symbols) and the exact solution (lines). 

At small $t_1$, the second-order dual perturbation theory is indistinguishable from the exact solution. At larger inter-dimer hopping, deviations are clearly visible, implying the importance of higher order dual processes. Still, the second-order theory gives a good impression of the shape and magnitude of spatial correlations. 

The relative deviations between the dual perturbation theory and the exact solution are most pronounced in the imaginary part at $\nu_5$, but even in that case, the absolute deviation is small. Note that we have used the isolated dimer as the auxiliary impurity model here. It is likely that the results could be improved by using additional bath sites in the impurity model.

\subsection{Zeeman splitting}

\begin{figure}
 \includegraphics{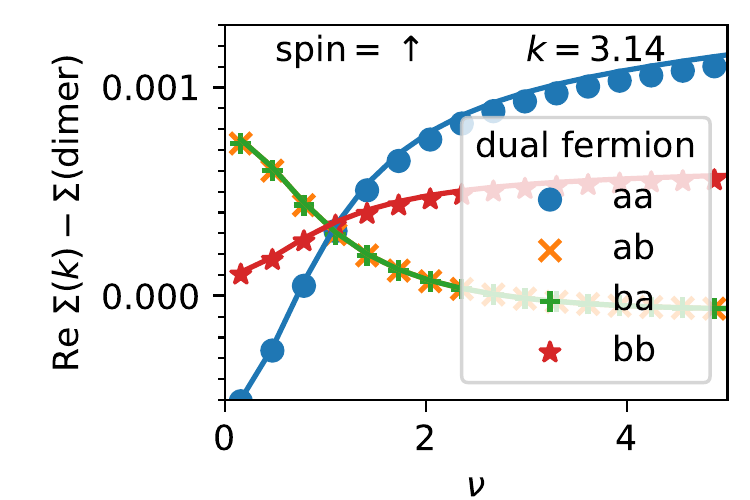}
 \includegraphics{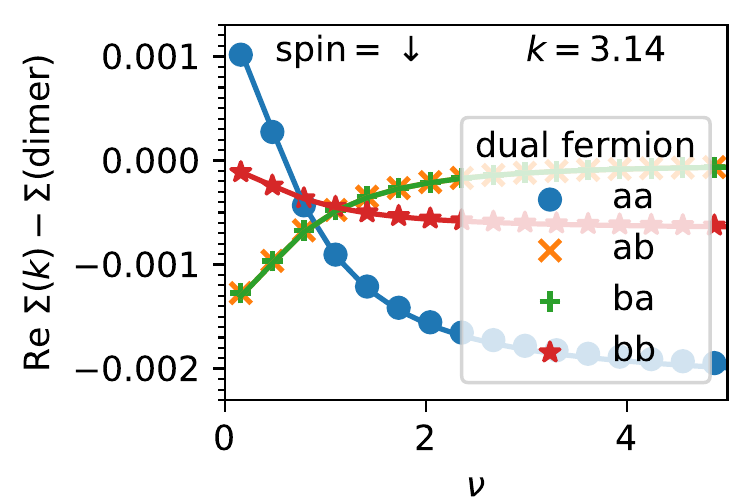} 
\caption{A Zeeman splitting of $B=0.1$ breaks the spin symmetry, leading to a different self-energy for $\up$ (top) and $\dn$ (bottom). This is a simulation of the four-site cluster with $t_1=0.01$, $t_0=4$, $U_a=10$, $U_b=3$, broken up into two dimers to do the dual perturbation theory. There is a good match between the dual perturbation theory (symbols) and the exact result (lines). }
\label{fig:zeeman}
\end{figure}

As an example of a system with non-equivalent blocks, we consider Zeeman splitting $B$ between $\up$ and $\dn$ spins,
\begin{align}
 H_\text{Zeeman} = - B \sum_{\text{sites }J} n_{J,\up}-n_{J,\dn}.
\end{align}
In this model, the Green's function is still block-diagonal, but the values in the two blocks are different. 
In the dual fermion calculation, this splitting is included on the level of the dimer impurity model, the dual perturbation theory then starts from a Green's function and vertex with explicitly broken spin symmetry. Since the splitting is accounted for at the impurity level, even large Zeeman splittings can be dealt with accurately.
The resulting self-energy is shown in Fig.~\ref{fig:zeeman}. In addition to Zeeman splitting, the dual fermion approach can also be applied to other models with broken spin symmetry including the mass-imbalanced Hubbard model~\cite{Philipp17} and the Falicov-Kimball model~\cite{Antipov14,Ribic16,Astleithner20}.

\section{Conclusion}

We have presented the second-order dual fermion approach for multi-orbital systems.
Our implementation is based on the open source \triqs\ toolbox and is interfaced with its impurity solvers. 
This development opens the road towards the investigation of spatial correlations in a wide set of models and real materials.

As examples, we have applied the method to SrVO$_3$ and to several small test systems. In SrVO$_3$, spatial correlations are present and visible in the calculations, but the magnitude of the non-local variations of the self-energy (100 meV scale) is an order of magnitude smaller than the local self-energy (1 eV scale) which is already captured in DMFT. Our results for SrVO$_3$ are compatible with those of a similar extension of DMFT, namely the dynamical vertex approximation~\cite{Galler19}.

In exactly solvable model systems, we have found that the dual perturbation theory used here is exact to second order in both the inter-cell hopping and in the interaction strength. This serves as a stringent test on the implementation. As the dual fermion method interpolates between these two opposite limits~\cite{Rubtsov08}, the results stay somewhat close to the exact solution even at intermediate values of the parameters.

The implementation presented here can deal with broken spin symmetry and with general interactions within the auxiliary impurity model. 
For interactions beyond the auxiliary impurity model, the dual boson generalization of dual fermion can be used~\cite{Rubtsov12,vanLoon14}, but that is beyond the scope of this work.

\acknowledgments

The author would like to thank 
Malte Sch\"uler for help with the SrVO$_3$ calculations and for feedback on the manuscript and 
Tim Wehling, Hartmut Hafermann, Alexander Lichtenstein, Friedrich Krien, Manuel Zingl, Anna Galler, Nils Wentzell and Igor Krivenko for useful discussions and advice.
This research is support by the Central Research Development Fund of the University of Bremen.
The North-German Supercomputing Alliance (HLRN) provided HPC resources (project hbp00047) that contributed to this work.

\bibliography{references}

\end{document}